\documentclass{vgtc}                          

\ifpdf
  \pdfoutput=1\relax                   
  \usepackage{graphicx}                
  \DeclareGraphicsExtensions{.pdf,.png,.jpg,.jpeg} 
\else
  \usepackage{graphicx}                
  \DeclareGraphicsExtensions{.eps}     
\fi%

\graphicspath{{figures/}{pictures/}{images/}{./}} 

\usepackage{microtype}                 
\PassOptionsToPackage{warn}{textcomp}  
\usepackage{textcomp}                  
\usepackage{mathptmx}                  
\usepackage{times}                     
\usepackage{cite}                      
\usepackage{tabu}                      
\usepackage{booktabs}                  

\usepackage{color}
\usepackage{xcolor}
\usepackage{url}
\usepackage{enumitem}
\usepackage{multirow}
\usepackage{wrapfig}

\usepackage{graphicx}
\usepackage{caption}
\usepackage{ulem}
\usepackage{xspace}

\onlineid{0}

\vgtccategory{Research}

\vgtcinsertpkg

\title{An Exploration And Validation of Visual Factors in Understanding Classification Rule Sets}

\author{Jun Yuan\thanks{e-mail: junyuan@nyu.edu} %
\and Oded Nov\thanks{e-mail:onov@nyu.edu} %
\and Enrico Bertini\thanks{e-mail:enrico.bertini@nyu.edu}}
\affiliation{New York University}

\abstract{Rule sets are often used in Machine Learning (ML) as a way to communicate the model logic in settings where transparency and intelligibility are necessary. Rule sets are typically presented as a text-based list of logical statements (rules). Surprisingly, to date there has been limited work on exploring visual alternatives for presenting rules. In this paper, we explore the idea of designing alternative representations of rules, focusing on a number of visual factors we believe have a positive impact on rule readability and understanding. We then presents a user study exploring their impact. The results show that some design factors have a strong impact on how efficiently readers can process the rules while having minimal impact on accuracy. This work can help practitioners employ more effective solutions when using rules as a communication strategy to understand ML models.%
} 

\CCScatlist{
  \CCScatTwelve{Human-centered computing}{Visualization}{Empirical studies in visualization}{}{};
}

\begin{document}


\firstsection{Introduction}

\maketitle

Rules have been used extensively in Machine Learning (ML) as an ``interface'' to communicate information about the logic used by a model (either for descriptive or predictive purposes). Many different types of rules exist in ML which capture different types of patterns. In this work we focus exclusively on \textit{classification rules} that describe associations between the feature values found in the data and an outcome of interest (e.g., health indicators and risk of diabetes). \looseness=-1

Rule-based machine learning (ML) models have been used for decades in decision-making and prediction~\cite{clark1991rule,liu1998integrating,evans2018learning,lakkaraju2016interpretable,letham2015interpretable,safavian1991survey,wang2017bayesian,yang2017scalable}. Although complex models such as DNN and ensemble models usually have higher performance, they are also limited by their complexity as well as low transparency and intelligibility; making their use limited in settings where transparency and user trust are important (e.g., healthcare, security, etc.). In recent years, the increasing need for model interpretability and explanation has led to a resurgence of rule-based models; with algorithms that aim at achieving similar levels of predictive power as less transparent solutions~\cite{lakkaraju2016interpretable, wang2015falling,wang2017bayesian}. Rules are also being used as a way to describe the behavior of existing black-box models by using model-agnostic explanation~\cite{guidotti2018local,sanchez2015towards,lakkaraju2019faithful,ribeiro2016should,ribeiro2018anchors}. 

To be interpretable, just being rule-based is not enough. Lakkaraju \textit{et. al}~\cite{lakkaraju2016interpretable} pointed out that rule lists (if-then-else structure) are more interpretable than a general decision tree (hierarchical structure) because of its reduced complexity and rule sets (if-then structure) are more interpretable than rule lists because people can describe decision boundaries more precisely. Although more and more rule-based approaches are introduced, rules are generally shown as logical statement expressed with text once they are generated. These rules can easily become long and complicated and, as such, hard to read and understand. An open question in this area is whether adopting alternative visual representations may lead to improvements in terms of rule sets readability and understanding. \looseness=-1

While initial attempts to produce more effective visual representations of rules have been made~\cite{di2019surrogate,kim2017firewall,mansmann2012visual,ming2019rulematrix,soares2020explaining}, progress in this area has been surprisingly limited in scope and not sufficiently systematic. More specifically, while prior work proposes alternative ways to visualize rules, we are not aware of systematic analyses of how rules \textit{could} be visualized and what factors may play a role in their effectiveness. \looseness=-1

In this work, we first propose an initial set of \textit{visual factors}. Then we focus on two specific factors: \textit{feature alignment} and \textit{predicate encoding} and study them through a crowdsourced controlled experiment with 338 participants. In all the tested tasks, the combination of aligned layout and graphical encoding reaches the best time performance without sacrificing accuracy and subjective confidence. These results contribute to how visual factors can support rule understanding and lead to new solutions of rule logic visualization. \looseness=-1
\section{Related Work}
\label{sec:background}

\begin{figure}
    \centering
    \includegraphics[width=0.44\textwidth]{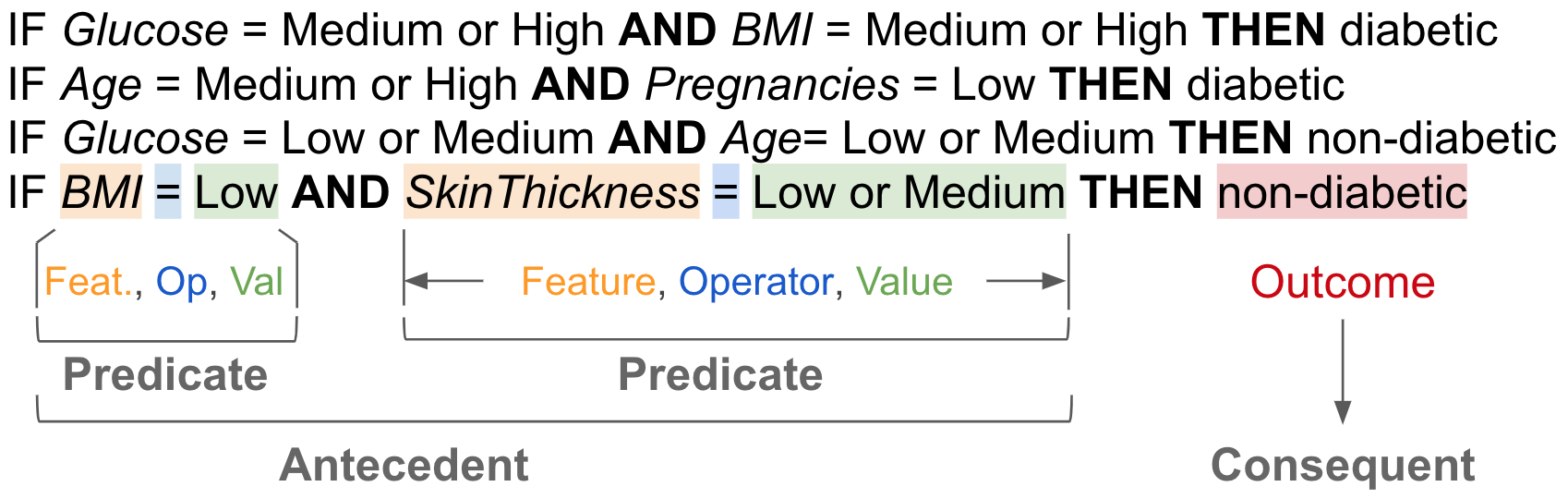}
    \vspace*{-.3cm}
    \caption{Terminology used to describe rules. Using rule set generated on the diabetes data set as an example.\looseness=-1}
    \label{fig:pima}
    \vspace*{-.5cm}
\end{figure}

\begin{figure*}
    \centering
    \includegraphics[width=.9\textwidth]{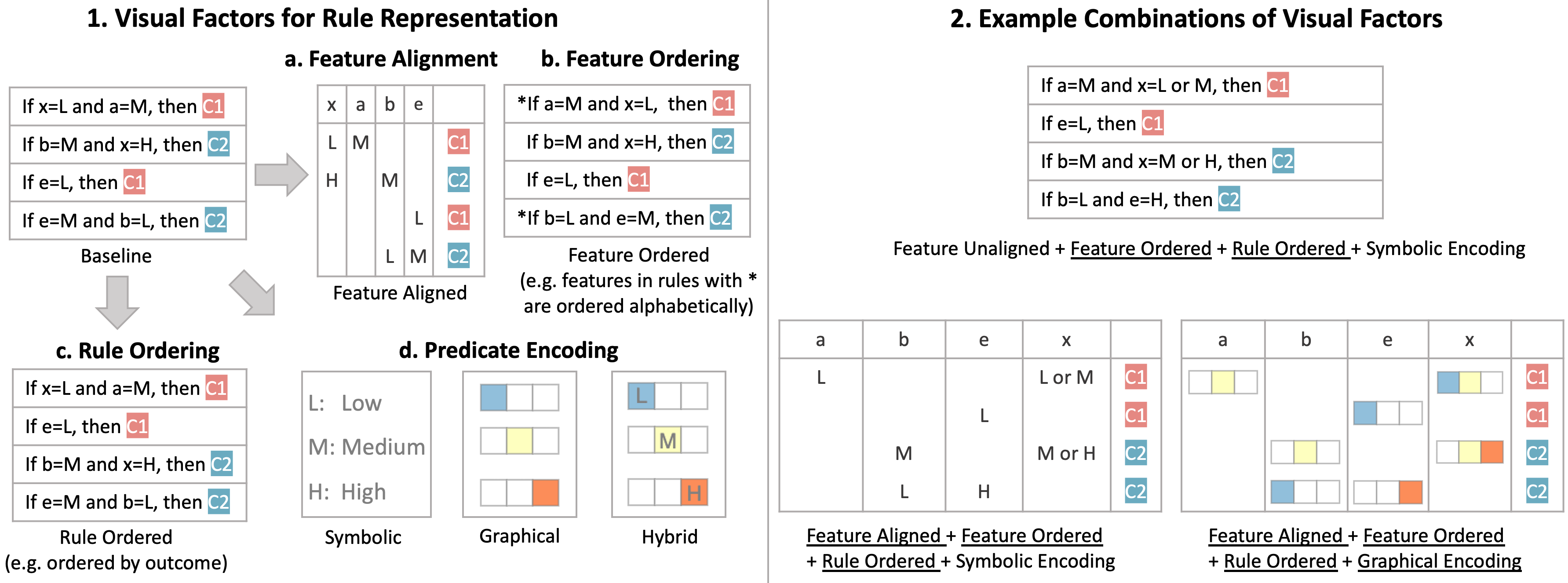}
    \vspace*{-.2cm}
    \caption{Part (1) includes four visual factors we identified that play a role in the process of human understanding a set of rules: a. Feature Alignment, b. Feature Ordering, and c. Rule Ordering which are strategies relevant to the \textit{Spatial Arrangement} of predicates; d. \textit{Predicate Encoding} includes the strategies of how to express the predicates in rules. Part (2) includes example combinations of different visual factors. \looseness=-1}
    \vspace*{-.5cm} 
    \label{fig:visual_factor}
\end{figure*}

Rules are structured as logical statements such as those presented in Figure~\ref{fig:pima}. 
Rule intelligibility can be expected to degrade as rule complexity increases. Recently researchers identified a number of factors that influence complexity with different names~\cite{lage2019evaluation,lakkaraju2016interpretable, wang2017bayesian}. Some of the most common include: the total number of attributes used in the predicates contained in the rule list (dimensionality), the total number of rules (cardinality), the amount of overlap between rule-related subgroups (redundancy), the maximum number of predicates that a rule can include (maximum rule length), the number of distinct values each predicate can have (feature cardinality).

While a complete investigation of how these factors impact rule understanding does not exist yet, some few empirical studies on rule structure exist. For example, empirical studies have been done to explore how human-interpretability and decision-making process can be influenced by different rule structures ~\cite{lakkaraju2016interpretable, subramanian1992comparison}, different levels of complexity~\cite{lage2019evaluation}, or both~\cite{huysmans2011empirical}. However, our focus is how visual representation may impact understanding of classification rules while keeping this information a fixed factor. 

\section{Visual Factors for Representation of Rule Sets}
\label{section:factors}
We now more systematically explore visual factors that can be used to design alternative visualizations of rule sets. In doing that, it is important to clarify the scope of our work in suggesting these factors. We focus exclusively on flat representations because not all rule algorithms produce hierarchical structures. Conversely, hierarchical rules can always be transformed into a list of rules.  

Visual factors for rules can be grouped into two main classes: \textbf{Spatial Arrangement} (how to position the rule components in the visual space, including the alignment of features and ordering of rule components) and \textbf{Predicate Encoding} (how to create visual representations of the values and conditions expressed in the rule's predicates). \looseness=-1

\textbf{Feature Alignment.} In a standard textual representation the features referenced in each predicate are \textit{unaligned}, that is, they have a different location. As shown in Figure~\ref{fig:visual_factor}-1:baseline, predicates using the same features (x, a, b, e) may be located in different horizontal positions. An alternative arrangement is to have the predicates \textit{aligned} into a tabular format as shown in Figure~\ref{fig:visual_factor}-1a. The unaligned arrangement is more compact but the aligned arrangement affords much easier comparison across the rules. \looseness=-1

\textbf{Feature Ordering.} Features and their corresponding predicates can also be ordered according to different criteria (e.g., alphabetically) (see Figure~\ref{fig:visual_factor}-1b).  \looseness=-1

\textbf{Rule Ordering.} Rules can also be ordered according to relevant criteria. In the mock-up we show in Figure~\ref{fig:visual_factor}-1c, the rules can be ordered based on the consequent/outcome part of the rule. 

\textbf{Predicate/Outcome Encoding.} The predicates of a rule are logical statements over a set of features, each with an associated domain of values. They can be connected by logic $AND$ or $OR$ connectives.
The outcome of a classification rule is also a predicate, typically with a single value of categorical or ordinal type. When confronted with the problem of visualizing predicates we have identified three main broad strategies as shown in Figure~\ref{fig:visual_factor}-1d: (1) \textit{Symbolic encoding} uses a mix of logical and textual symbols (this is the most common solution found in papers and software packages); (2) \textit{Graphical encoding} uses graphical representations rather than symbols. We propose a design that uses both position and color encoding to represent the values of the predicates; (3) \textit{Hybrid encoding} uses a mix of the two strategies above by integrating symbols and graphical encoding in one solution. We created a website\footnote{\url{https://rule-logic-vis.herokuapp.com/}} to present the rule visualization of all the visual factors we identified using rules at different complexity levels (note that the solutions presented in Figure~\ref{fig:visual_factor} are mock-ups, whereas those presented in the website correspond to those used in the study). \looseness=-1

In light of this characterization, the issue we address in this study, regarding how to represent the predicates, is how to balance the benefits of symbolic encoding and those of graphical encoding. In theory, symbolic encoding should be slower and more effortful than graphical encoding, especially as the number of rules increases. However, the unfamiliar nature of graphical encoding may slow down viewers due to increased uncertainty, less precision and lack of familiarity with the medium. To start addressing this question, we present an empirical study in Section~\ref{section:experiment} to gather evidence about how these strategies compare when used to perform complex tasks with lists of rules. \looseness=-1


\section{Experiment}
\label{section:experiment}
To better understand the impact of the identified visual factors, we designed a controlled experiment that aims at teasing out their effect on rule understanding. In this section we describe the design of the experiment and report the results.

\subsection{Experiment Design}

\subsubsection{Stimuli: Visual Representations and Rule Sets.}
As described in Section~\ref{section:factors}, the ordering strategies can be applied to any given visual representation. Different ordering can be utilized to prioritize information that humans need for different purposes. So we do not test them in this study. Instead, we fixed the order of the rules and decided to sort by outcome. Similarly, we fixed the order of the features and opted to sort by frequency of features in the rule set; with frequency decreasing from left to right. The study we present is organized as a controlled experiment in which we compare conditions of the alignment strategy (aligned, unaligned) and the predicate encoding strategy (symbolic, graphical, hybrid).  

The main goal of the study is to understand the impact our interventions have with respect to the traditional representation of rules based on text. For this reason our study uses a standard textual list of rules as a \textit{control condition} and organizes the study around improvements over it. We include the following four conditions:

\begin{itemize}[noitemsep,topsep=1pt]
    \item[(\texttt{SU})] \textbf{S}ymbolic encoding for predicates, \textbf{U}naligned features (baseline, control condition);
    \item[(\texttt{SA})] \textbf{S}ymbolic encoding for predicates, \textbf{A}ligned features;
    \item[(\texttt{GA})] \textbf{G}raphical encoding for predicates, \textbf{A}ligned features;
    \item[(\texttt{HA})] \textbf{H}ybrid encoding (graphical and symbolic) for predicates, \textbf{A}ligned features.
\end{itemize}

Having to choose a specific data type to cover, we opted for ordinal data (e.g., Low, Medium and High as feature values) since it covers useful properties of both categorical and quantitative data.
 \looseness=-1

For the purpose of this study we use two main data sets: a credit risk data set~\cite{fico} in the training phase to help the participants to familiarize with the tasks and concepts of the study; and a diabetes data set~\cite{smith1988using} for the actual study.\looseness=-1

To generate the rules, we used the algorithm proposed by Wang \textit{et al.}~\cite{wang2017bayesian}. For training, we generated a set of $6$ rules over $5$ features. As for testing, we created a rule set with an higher level of complexity ($18$ rules over $8$ features) in order to differentiate the effects of the tested visual factors. Each rule in the test set has a maximum of $3$ predicates in the antecedent. In a recent work~\cite{lakkaraju2016interpretable}, $9$ rules are considered for an evaluation of rule understanding. We also found in our pilot study that the size of our test rule set is neither too simple nor too complex for humans to understand. \looseness=-1

In designing the rules we also had to mitigate the effect of \textit{prior knowledge}. In one of our pilot studies, we tested the rules generated from a diabetes data set and found that the participants could predict the answer by using intuition and with little engagement with the visualization. To mitigate this effect, we changed the feature names (originally \textit{Age}, \textit{BMI}, etc.) into names of minerals (\textit{Magnesium}, \textit{Calcium}, etc.). The subjects were instructed to analyze the predictions of a model that used the mineral features to predict the level of risk of contracting a hypothetical disease.  \looseness=-1

\subsubsection{Tasks} To simulate the process of humans understanding rule logic, we designed two sets of test tasks: one set for \textbf{Prediction Estimation (T1)} and one for \textbf{Prediction Characterization (T2)}. For T1 we asked the participants to answer questions like: \textit{``What is the most common prediction for rules containing conditions that match a person with a High value of \textit{Calcium}?''}. For T2 we asked questions like: \textit{``Considering only the rules that predict high risk, what is the most common value for Magnesium?''}. For both types of questions increased complexity progressively.
With variations in complexity we created a total of $10$ test questions: $5$ for Task 1 and $5$ for Task 2. More details can be found in the Github page \footnote{\url{https://github.com/junyuanjun/rule_empirical_study}}. \looseness=-1

\subsubsection{Procedure.}
We presented the 4 conditions outlined above to 4 separate groups of participants using a between-subjects design. The study was organized according to the following steps which were common to every group: (1) We started with a consent form and a collection of \textit{demographic information}; (2) We asked the participants to watch a 4-minute \textit{tutorial} video of basic ML concepts and how to read the rule visualizations, which is followed by a verification quiz. This was used as a pre-requisite to move on; (3) We described the tasks through a \textit{task introduction} page showing example tasks and asked them to perform a few simplified tasks resembling those used in the test; (4) In the \textit{test} stage, we first showed the visualization used in the test and described the test process. Then, for each task we asked the participants to answer 6 questions for each task type. Each question was posed as a choice in a multiple-choice test where only one out of three answers was right. For each question the participant also had to specify the confidence level associated to the answer using 5-point likert scale with $1$ being 'not confident at all' and $5$ being 'extremely confident'. The first question of each task type was used only for habituation to the task and was eliminated from the analysis. In the end, the participants had to report their overall subjective sense of effort in performing the tasks. 




\begin{table*}
    \centering
    \begin{tabular}{l| l l l | l l l| l l l | l l l }
        \hline
          \multirow{2}{*}{\textbf{Task}} & \multicolumn{3}{c|}{\textbf{Accuracy (0.00-1.00)}} & \multicolumn{3}{c|}{\textbf{Response Time (s)}} & \multicolumn{3}{c|}{\textbf{Confidence (1-5)}} & \multicolumn{3}{c}{\textbf{Subjective Effort (1-5)}}\\
          & \texttt{SA}-\texttt{SU} & \texttt{GA}-\texttt{SU} & \texttt{HA}-\texttt{SU} & \texttt{SA}-\texttt{SU} & \texttt{GA}-\texttt{SU} & \texttt{HA}-\texttt{SU} & \texttt{SA}-\texttt{SU} & \texttt{GA-SU} & \texttt{HA-SU} & \texttt{SA-SU} & \texttt{GA-SU} & \texttt{HA-SU}\\
        \hline
      \textbf{T1}  & 0.03 & 0.02 & 0.02 & -18.40 & -19.35 & -18.63 & 0.25 & 0.21 & 0.18 & \multirow{2}{*}{-0.27} & \multirow{2}{*}{-0.03} & \multirow{2}{*}{-0.10}\\
      \textbf{T2}  & 0.00 & -0.02 & 0.01 & -7.44 & -11.49 & -9.90 & 0.12 & 0.05 & -0.08 \\
        \hline
    \end{tabular}
    \vspace*{-.3cm}
    \caption{Absolute effect size for the comparison with the control condition (\texttt{SU}).}\vspace*{-.5cm}
    \label{table:cohens_d}
\end{table*}

\begin{figure}
    \centering
    \includegraphics[width=.45\textwidth]{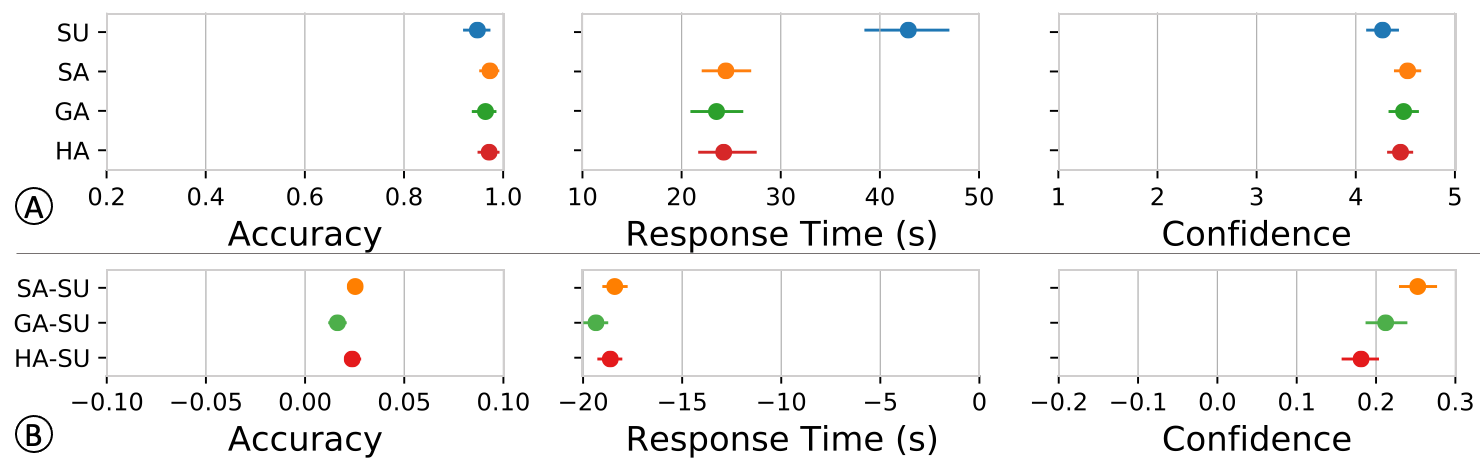}
    \vspace*{-.2cm}
    \caption{Performance of Task 1. (A) shows the mean values and $95\%$ CI of accuracy, time, and confidence. (B) shows the absolute effect size as well as the confidence interval of the comparison between aligned conditions (\texttt{SA}, \texttt{GA}, \texttt{HA}) with the baseline (\texttt{SU}).}
    \vspace*{-.5cm}
    \label{fig:task1}
\end{figure}

\subsubsection{Performance Metrics}
To evaluate the performance of rule reading tasks, we used the following four metrics:
\begin{itemize}[noitemsep,topsep=1pt]
    \item \textit{Accuracy:} The number of correct answers over the total number of questions.
    \item \textit{Response Time (RT):} The time required to answer a question.
    \item \textit{Confidence:} The subjective confidence scores for each question from 1 to 5 (1: ``not confident at all'', 5: ``extremely confident'').
    \item \textit{Subjective Effort:} The subjective sense of effort scores collected in the end of the study (value ranging between 1-5, 1 being high effort and 5 being low effort).
\end{itemize}

\subsubsection{Participants.}
We recruited our participants through the crowdsourcing platform Prolific. Our population sample consisted of 338 people from either the US or UK, with an approval rate of at least $99\%$. We paid each subject a total of \$4 USD for a test duration of about $25$min.




\subsubsection{Hypotheses.}
We developed the following hypotheses for the study. All the hypotheses, as well as the experimental design and the pre-planned analysis, can be found in our pre-registered study \footnote{\url{https://osf.io/79ujk}}.


\begin{itemize}[noitemsep,topsep=.5pt]
    \item[\textbf{H1}] Response Time (RT): We expected alignment to have the strongest impact on speed. Therefore we expected unaligned representation (\texttt{SU}) to take substantially more time than feature-aligned representations (\texttt{SA}, \texttt{GA}, \texttt{HA}). As for the comparison between encodings we expected \texttt{HA} to have an advantage over \texttt{GA} and \texttt{HA} because it enables both symbolic and graphical reading of the conditions. We expected these differences to be less pronounced than the difference afforded by alignment. \looseness=-1
    \item[\textbf{H2}] Accuracy: Our tasks are designed to focus more on efficiency than correctness. We did not expect any substantial differences in accuracy between the conditions.
    \item[\textbf{H3}] Confidence: We expected the control condition (\texttt{SU}) to have higher confidence scores due to the level of familiarity our subjects may have with this solution compared to the more unfamiliar ones we used in the other conditions. 
    \item[\textbf{H4}] Subjective Effort: We expected the participants to assign to the baseline (\texttt{SU}) way higher effort scores than to the other conditions. We did not have specific expectations for \texttt{SA}, \texttt{GA} and \texttt{HA}. 
\end{itemize}

\subsection{Results}
The dropout rate for those who did not finish the study was $10\%$. In the complete replies we collected, we excluded the subjects who performed the tasks too fast ($<5$s per question) or too slow ($>100s$ per question for T1 or $>120s$ per question for T2) as stated in the pre-registration. Then we conducted the analysis based on these samples (n=73, 78, 75, 77 for \texttt{SU}, \texttt{SA}, \texttt{GA}, \texttt{HA}). 

All results are presented using absolute effect size and bootstrap $95\%$ confidence interval as suggested in~\cite{dragicevic2016fair}. We on purpose avoid the calculation of p-values to avoid dichotomous thinking~\cite{besanccon2019continued}. We use in their place the language of estimation driven by effect sizes and confidence intervals as suggested by Cumming in~\cite{cumming2013understanding}. For the analysis of the results, we use \texttt{SU} as a baseline, that is, the results of the other conditions will be presented in relation to the control case. 
An overview of the results is shown in Table~\ref{table:cohens_d}. \looseness=-1

\subsubsection{Task 1: Prediction Estimation}
We provide the performance of Task 1 across visual conditions in Figure ~\ref{fig:task1}. Accuracy across all conditions is very similar and very close to $100\%$. And all the aligned conditions have extremely similar accuracy values with highly negligible differences. 

In terms of the efficiency, the response time with the control condition is substantially higher. The average response time for performing Task 1 with (\texttt{SU}) is over $40s$, while the response time with the other aligned conditions (\texttt{SA}, \texttt{GA}, \texttt{HA}) is less than $30s$. For this task, we observe that alignment has a stronger influence than encoding for that all the feature-aligned conditions result in similar amount of time reduction. As shown in Figure~\ref{fig:task1}B:Response Time, the participants with feature-aligned conditions (\texttt{SA}, \texttt{GA}, \texttt{HA}) spent around $19s$ less for Task 1 on average than the participants reading rules without feature alignment (\texttt{SU}). This is a large efficiency improvement considering that users with the baseline condition spend around $40s$ on average. When comparing the encoding strategies over the alignment (\texttt{SA}, \texttt{GA}, \texttt{HA}) we can see that the differences are way less pronounced. Out of the 3 conditions the Graphical encoding (\texttt{GA}) seems to have a slight advantage of $0.95s$ faster than Symbolic encoding (\texttt{SA}) and $0.72s$ faster than Hybrid encoding (\texttt{HA}).  \looseness=-1

Alignment also leads to an improvement in subjective confidence. Symbolic encoding brings the most improvement around $6.25\%$ (0.25/4), while graphical encoding takes the second place around $5\%$ (0.2/4), hybrid encoding around $4.5\%$ (0.18/4).  \looseness=-1


\subsubsection{Task 2: Prediction Characterization}
\begin{figure}
    \centering
    \includegraphics[width=.45\textwidth]{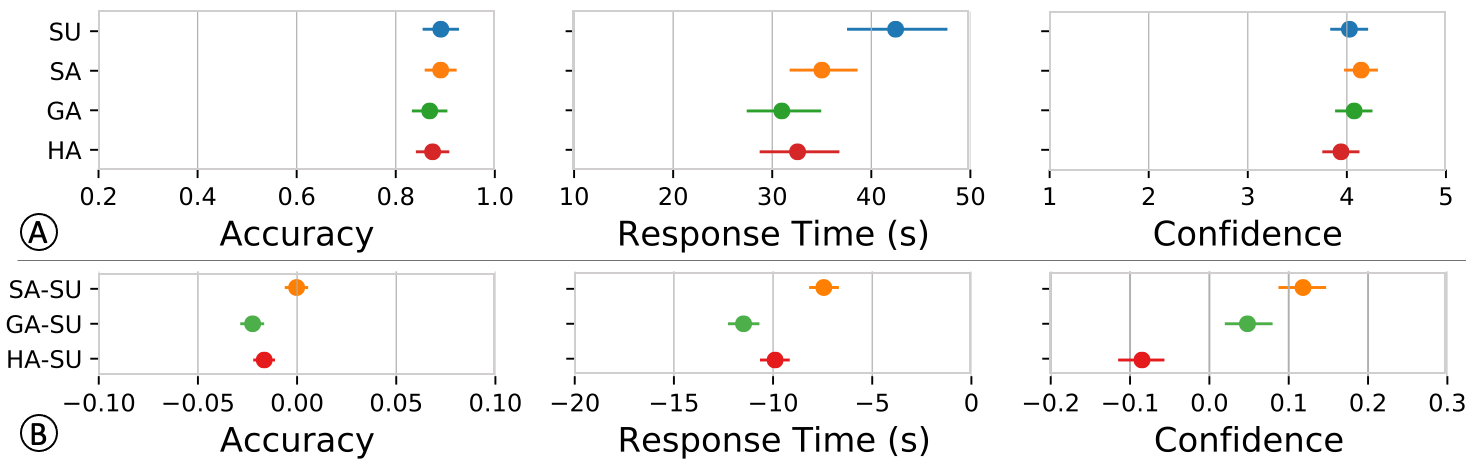}
    \vspace*{-.3cm}
    \caption{Task 2 Performance. (A) shows the mean values and $95\%$ CI for each performance metric. (B) includes the absolute effect size and $95\%$ CI compared with \texttt{SU}.}  
    \vspace*{-.45cm}
    \label{fig:task2}
\end{figure}

In Figure ~\ref{fig:task2}, we provide the performance of Task 2 based on three measurements. Similarly, participants with all conditions reached a relatively high accuracy. The influence of visual encoding and feature alignment on accuracy is more pronounced in comparison to Task 2. The three feature aligned conditions have a similar level of accuracy as the control condition.

We also observe the efficiency improvement results from feature alignment and predicate encoding. It is clear that participants who perform Task 2 with feature-aligned rules require at least $5s$ less than with the baseline representation. Comparing the encoding strategies along with alignment, rules contain graphical encoding (\texttt{HA} and \texttt{GA}) lead to larger improvement in response time. More specifically, compared with the average response time from baseline (around $40s$), the aligned feature along with graphical encoding reduces the most, around $11.49s$; hybrid encoding (\texttt{HA}) results in $9.90s$ less; symbolic encoding (\texttt{SA}) leads to around $7.44s$ less. \looseness=-1

As for confidence, the difference is slight. Symbolic encoding (\texttt{SA}) leads to the highest subjective confidence on average with around $3\%$ (0.12/4) higher than the confidence from the baseline. Then graphical encoding (\texttt{GA}) and hybrid encoding (\texttt{HA}) strategies lead to around $1.25\%$ (0.05/4) more and $2\%$ (0.08/4) less confidence score compared with the baseline.  \looseness=-1


In the end, subjects with the control condition (\texttt{SU)} report the highest subjective effort and subjects with other conditions report slightly less subjective effort. According to the overall subjective effort score,  \texttt{SA} leads the most improvement around $6.68\%$ (0.27/4), then \texttt{HA} $2.5\%$ (0.1/4), \texttt{GA} $0.75\%$ (0.03/4).

\section{Discussion}
Our results partially confirm some of our hypotheses. Alignment clearly has a strong impact on performance. All designs that use feature alignment have substantial improvements in response time, with no substantial effects on accuracy, confidence or workload. The effect on response time is very large: around $20s$ improvement with aligned designs on T1 and about $10s$ improvement for T2. Such a large difference on a single inference task can have a substantial effect when considering how many of these single inferences will be carried out to make sense of a set of rules in a real-world setting. \looseness=-1

Contrary to our expectations, the unaligned design did not lead to higher confidence. This may be due to the fact that performing the assigned tasks with the control condition requires more time than with the other ones. If an effect of familiarity exists it is probably cancelled out by workload (time). More research is needed to understand how workload and familiarity interact. 

The effect of predicate encoding is way less pronounced than the effect of alignment. Across the two tasks graphical encoding with alignment (\texttt{GA}) seems to have a slight advantage in terms of response time over the other aligned designs (\texttt{SA}, \texttt{HA}). Such an advantage is more pronounced in Task 2 where \texttt{GA} has about a mean $4s$ improvement over \texttt{SA} and about $1.5s$ over \texttt{HA}. 
Regarding why such an improvement exist with \texttt{GA}, we do not have a definite explanation. Our original intuition was that the hybrid design (\texttt{HA}) would integrate the benefits of symbolic and graphical representations, but our results do not confirm this intuition. Including a symbolic representation in a predicate seems to slow down performance, even if slightly, maybe due to more time spent reading the symbols. \looseness=-1


Extrapolating from a single study is always difficult and we believe more studies in this space will be needed to build a more accurate understanding of the effect of visual factors on rule understanding. However, we feel confident for the conclusions based on what we have observed. The effect of alignment is extremely strong and we expect this effect to be easy to replicate and robust to many variations (e.g., different ways to encode the predicates). The effect of graphical encoding is less pronounced but not necessarily negligible. 
Equipped with this knowledge we feel confident in suggesting the use of aligned layouts with graphical representations (\texttt{GA} and \texttt{HA}). They speed up performance considerably without affecting accuracy and with very negligible effects on confidence. The only downside we think designers should be aware of is that the aligned designs require more space to be presented. Since they use space for alignment they tend to be substantially less compact than the standard design where alignment is not present. \looseness=-1

\section{Conclusion And Future Work}
This work introduces a set of visual factors in terms of understanding classification rules. Our empirical study confirms the influence of two visual factors: feature alignment, and predicate encoding. The analysis of the study results reveal that feature alignment can improve the efficiency of rule understanding tasks while staying accurate. Moving forward, We are particularly interested in further investigating whether visual factors play a role in a more complex model interpretation process, and in the design of more effective visualizations to assist users to utilize rule information. \looseness=-1

\acknowledgments{We thank all the study participants and reviewers for their comments. This work was partially supported by a contract with Capital One and the NSF award number 1928614.}

\bibliographystyle{abbrv-doi}

\bibliography{template}
\end{document}